\documentclass[sigconf]{acmart}
\AtBeginDocument{%
  }

\acmSubmissionID{488}
\usepackage{graphicx}  
\usepackage{subcaption} 
\usepackage{ragged2e}
\usepackage{balance}
\copyrightyear{2025} 
\acmYear{2025} 
\setcopyright{acmlicensed}\acmConference[MM '25]{Proceedings of the 33rd
 ACM International Conference on Multimedia}{October 27--31, 2025}{Dublin,
 Ireland}
 \acmBooktitle{Proceedings of the 33rd ACM International Conference on
 Multimedia (MM '25), October 27--31, 2025, Dublin, Ireland}
 \acmDOI{10.1145/3746027.3754758}
 \acmISBN{979-8-4007-2035-2/2025/10}

\begin{document}

\title{DRKF: Decoupled Representations with Knowledge Fusion for Multimodal Emotion Recognition}

\author{Peiyuan Jiang}
\affiliation{%
  \institution{School of Computer Science and Engineering, University of Electronic Science and Technology of China}
  \city{Chengdu}
  \state{Sichuan}
  \country{China}}
\email{darcy981020@gmail.com}

\author{Yao Liu}
\authornotemark[1]
\affiliation{%
  \institution{School of Information and Software Engineering, University of Electronic Science and Technology of China}
  \city{Chengdu}
  \state{Sichuan}
  \country{China}}
\email{liuyao@uestc.edu.cn}

\author{Qiao Liu}

\affiliation{%
  \institution{School of Computer Science and Engineering, University of Electronic Science and Technology of China}
  \city{Chengdu}
  \state{Sichuan}
  \country{China}}
\email{qliu@uestc.edu.cn}

\author{Zongshun Zhang}
\affiliation{%
  \institution{School of Computer Science and Engineering, University of Electronic Science and Technology of China}
  \city{Chengdu}
  \state{Sichuan}
  \country{China}}
\email{202421081411@std.uestc.edu.cn}

\author{Jiaye Yang}

\affiliation{%
  \institution{School of Computer Science and Engineering, University of Electronic Science and Technology of China}
  \city{Chengdu}
  \state{Sichuan}
  \country{China}}
\email{202411081710@std.uestc.edu.cn}

\author{Lu Liu}

\affiliation{%
  \institution{School of Computer Science and Engineering, University of Electronic Science and Technology of China}
  \city{Chengdu}
  \state{Sichuan}
  \country{China}}
\email{202522080827@std.uestc.edu.cn}

\author{Daibing Yao}

\affiliation{%
  \institution{Yizhou Prison, Sichuan Province}
  \city{Chengdu}
  \state{Sichuan}
  \country{China}}
\email{357497551@qq.com}

\thanks{*Corresponding author: Yao Liu}

\renewcommand{\shortauthors}{Peiyuan Jiang, Yao Liu, Qiao Liu, Zongshun Zhang, Jiaye Yang, Lu Liu, \& Daibing Yao.}

\begin{abstract}
Multimodal emotion recognition (MER) aims to identify emotional states by integrating and analyzing information from multiple modalities. However, inherent modality heterogeneity and inconsistencies in emotional cues remain key challenges that hinder performance. To address these issues, we propose a Decoupled Representations with Knowledge Fusion (DRKF) method for MER. DRKF consists of two main modules: an Optimized Representation Learning (ORL) Module and a Knowledge Fusion (KF) Module. ORL employs a contrastive mutual information estimation method with progressive modality augmentation to decouple task-relevant shared representations and modality-specific features while mitigating modality heterogeneity. KF includes a lightweight self-attention-based Fusion Encoder (FE) that identifies the dominant modality and integrates emotional information from other modalities to enhance the fused representation. To handle potential errors from incorrect dominant modality selection under emotionally inconsistent conditions, we introduce an Emotion Discrimination Submodule (ED), which enforces the fused representation to retain discriminative cues of emotional inconsistency. This ensures that even if the FE selects an inappropriate dominant modality, the Emotion Classification Submodule (EC) can still make accurate predictions by leveraging preserved inconsistency information. Experiments show that DRKF achieves state-of-the-art (SOTA) performance on IEMOCAP, MELD, and M3ED. The source code is publicly available at \url{https://github.com/PANPANKK/DRKF}.
\end{abstract}


\maketitle

\section{Introduction}
Multimodal emotion recognition based on speech and text is essential for human-computer interaction (HCI) \cite{article2}. 
The MER's fundamental concept is to acquire modality representations and subsequently fuse them \cite{article5,article6}. In representation learning, contrastive learning-based methods have been widely applied to various multimodal tasks \cite{articlenew12,jenni2023audio,akbari2021vatt}. These methods rely on the multi-view redundancy assumption, which states that the shared information across different modalities can sufficiently capture the critical features required for downstream tasks \cite{articlenew9,articlenew10}.

\begin{figure}[ht]
    \centering
    \includegraphics[width=0.46\textwidth]{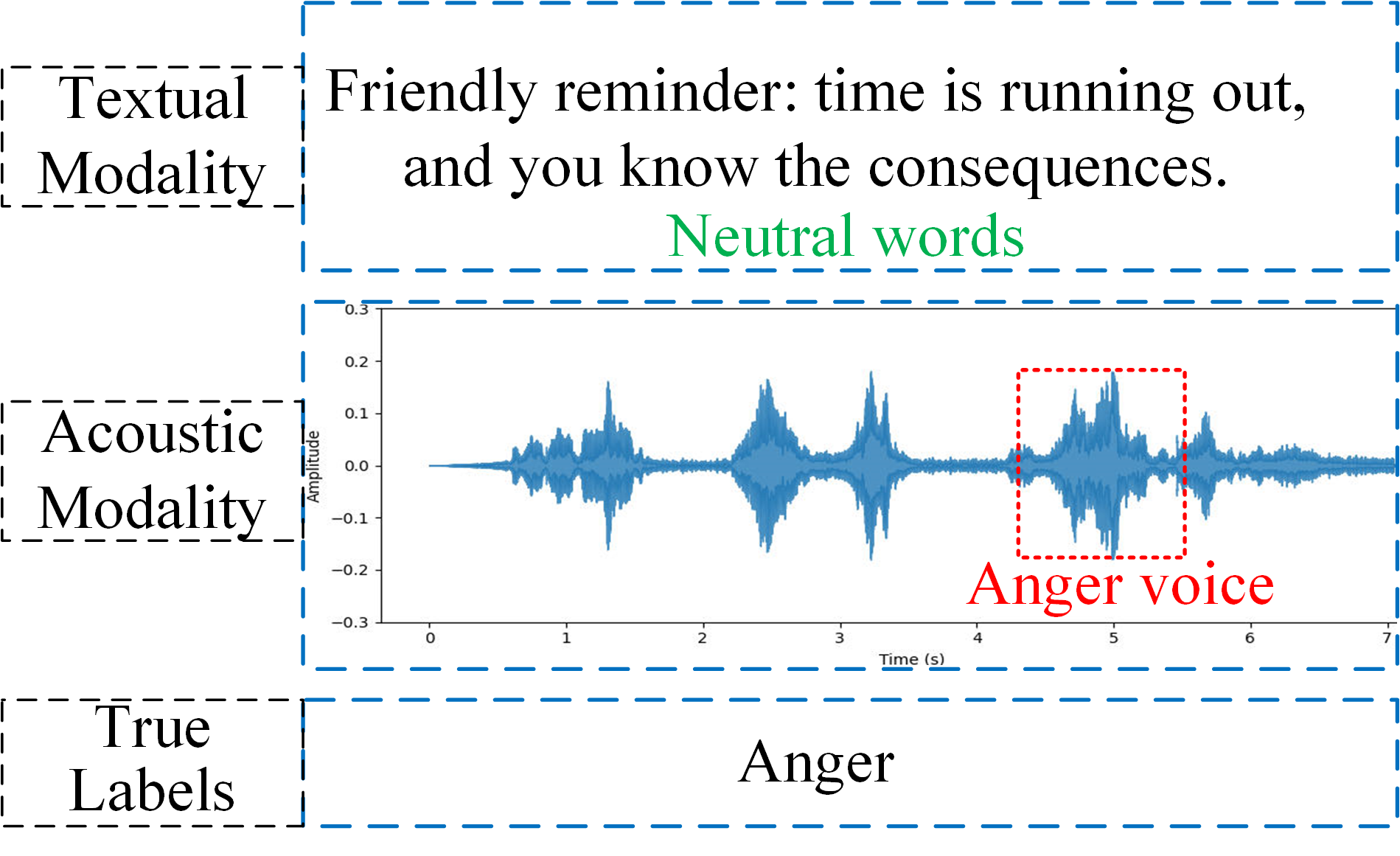}
    \caption{Illustration of Modality-specific Emotions and True Labels in a Multi-view Non-redundant Scenario.}
    \Description{Semantic conflict example: Text shows neutrality, audio detects anger, 
    yet the true label is anger (contradicting the textual modality).}
    \label{fig:fig1}
\end{figure}

Although multimodal emotion representation learning has made significant progress under this assumption, it does not always hold in broader real-world multimodal scenarios. In multi-view non-redundant scenarios, the information contained in each modality is not necessarily relevant to the downstream task. To address this issue, existing studies \cite{article8,articlenew15,article9,articlenew16} leverage techniques such as adversarial learning, parameter sharing, and subspace learning to decouple modality-specific and shared features.

The aforementioned methods are capable of extracting modality-specific and shared features. However, they are unable to ensure that the learned representations are pertinent to the tasks. \cite{shen2024beyond} adopts an information-theoretic perspective to describe the modality-specific and shared information pertaining to a given task. A Contrastive Mutual Information Estimation (CMIE) method has been introduced to optimize modality-specific and shared representations that are applicable to the task \cite{article41}. In such methods, neural networks (NNs) are typically employed to determine mutual information scores between modalities \cite{caron2020unsupervised, Bachman19Learning}. However, the efficacy of representation learning can be compromised by the instability of these approaches when there is a substantial distributional disparity between task labels and input modalities.

In multimodal representation fusion, early methods primarily relied on tensor fusion and simple feature concatenation \cite{tensorfusion, articlenew20}. Recent advances have introduced cross-attention mechanisms to better model semantic dependencies and task-oriented alignment across modalities. However, these mechanisms can still suffer from the introduction of redundant noise and increased modeling ambiguity, particularly when emotion-related information is inconsistently conveyed across different modalities \cite{articlenew22, articlenew23}.

In this work, we aim to address two key challenges in MER: the difficulty of extracting and aligning task-relevant information across heterogeneous modalities—stemming from their inherent representational differences—and the inconsistencies in emotional cues conveyed by different modalities, as illustrated in Fig.~\ref{fig:fig1}. To tackle these challenges, we propose a Decoupled Representations with Knowledge Fusion Method (DRKF) for MER. Our model consists of two modules: \textit{the Optimized Representation Learning Module (ORL)}, inspired by \cite{Jiang2023UnderstandingAC}, indirectly aligns the input modalities with the label distribution through progressive modality augmentation learning, thereby overcoming the challenge of mutual information estimation in the CMIE method caused by modality-label distribution discrepancies. Following the ORL, the Knowledge Fusion Module (KF) consists of a Fusion Encoder (FE), an Emotion Classification Submodule (EC), and an Emotion Discrimination Submodule (ED). The FE, based on self-attention mechanism, identifies the dominant modality of the current sample and integrates complementary emotional information from other modalities to enhance the fused representation. Under emotionally inconsistent conditions, the ED further constrains the fused representation to retain discriminative cues regarding intermodal emotional discrepancies, thereby mitigating potential errors caused by incorrect dominant modality selection. Finally, the EC takes the fused representation as input to perform the emotion classification task. Through their collaborative design, these three components enable more robust and adaptable multimodal emotion recognition.
\begin{itemize}
\item We introduce an optimized representation learning module, which learns an optimal enhanced modality to guide the alignment of distributions across modalities as well as between modalities and labels, thereby facilitating more effective representation decoupling.
\item We introduce a Knowledge Fusion Module that leverages collaborative learning to integrate fusion encoding, emotion consistency discrimination, and emotion classification, ensuring reliable emotion recognition.
\item Extensive experiments on three benchmark datasets demonstrated that the proposed DRKF framework surpasses state-of-the-art methods.
\end{itemize}

\begin{figure*}[ht]
    \centering
    \includegraphics[width=1.0\textwidth]{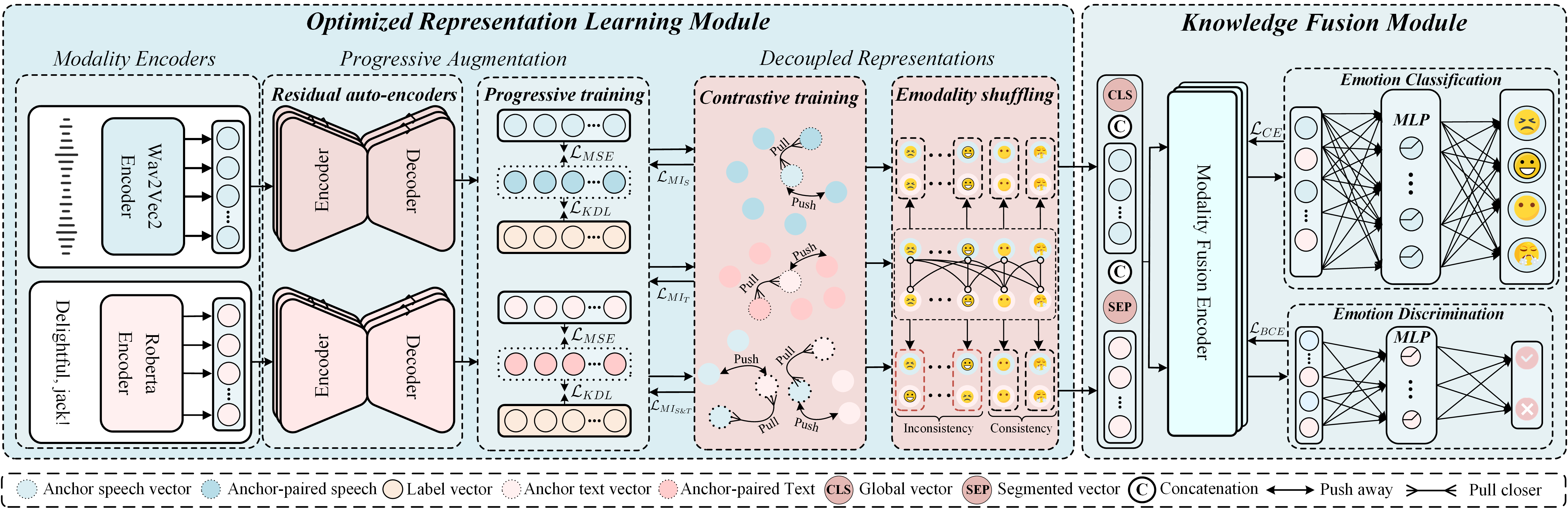}  
\caption{Overview of the Proposed DRKF Framework. It consists of two components: the ORL Module, which improves task-relevant modality mutual information estimation, and the KF Module, which models modality interactions for final emotion classification.}
\Description{Our proposed DRKF architecture integrates two complementary components: (1) the ORL module that optimizes mutual information extraction between task-relevant modalities, and (2) the KF module that learns hierarchical cross-modal relationships for robust emotion classification.}
    \label{fig:fig2}  
\end{figure*}

\section{Related Work}
\label{Related Work}
\subsection{Representation Learning}
Bengio \cite{article19} emphasized that the performance of machine learning models heavily depends on the selection of input features. Different feature representations can entangle varying underlying explanatory factors, potentially obscuring their distinct contributions to the learning process. Multimodal emotion representations can be categorized into feature engineering-based representations and deep neural network-based representations. The former relies on expert knowledge, including acoustic features extracted using openSMILE \cite{article21} and textual features from emotion lexicons \cite{article22} and syntactic structures \cite{article24}, while the latter leverages deep learning to automatically extract high-level features. Representative models include BERT \cite{article26}, RoBERTa \cite{article30}, wav2vec 2.0 \cite{article31}, and WavLM \cite{article33}.

Building on deep representations, multimodal emotion representation learning can be further categorized into single-stream and multi-stream models. Single-stream models use a shared encoder to learn joint representations within a unified latent space, whereas multi-stream models maintain separate pathways for each modality and integrate them later to capture cross-modal interactions \cite{lin2024adapt}. While multi-stream architectures effectively preserve modality-specific information and enhance cross-modal interactions, they can also introduce irrelevant noise, hindering optimal fusion performance. Decoupled representation serves as a key mechanism to filter out irrelevant information and improve fusion quality \cite{article8}.

To enhance the effectiveness of decoupled representations, recent studies have incorporated contrastive learning to reduce inter-modal distributional discrepancies and achieve decoupling of shared representations across modalities \cite{article35, article36}. Additionally, some approaches employ subspace mapping techniques, adversarial learning, and orthogonality constraints to extract task-relevant modality-specific information \cite{article16, articlenew15, article9}, while others leverage information-theoretic methods to quantify the task relevance of private information, further improving the interpretability and discriminative power of learned representations \cite{article41, article39, article40}.

\subsection{Modality Fusion}
Morency et al. \cite{article42} identified five core challenges in multimodal learning: representation learning, modality conversion, modality alignment, co-learning, and modality fusion. Among these, modality fusion is essential for cross-modal knowledge integration. Existing fusion strategies can be broadly categorized into feature-level, decision-level, and interaction-based fusion. Feature-level fusion combines features from different modalities, such as through concatenation \cite{article43} or time-scale-aware integration \cite{article44}, but increases the classifier's burden in handling redundancy and modality misalignment. Decision-level fusion \cite{zhang2023developing} integrates modality-specific predictions using methods like ensemble learning, weighted averaging, or voting, but often overlooks fine-grained interactions crucial for tasks like emotion recognition. Interaction-based fusion learns cross-modal relationships through attention mechanisms or latent space alignment. For instance, \cite{article45} proposed a multi-hop attention mechanism, allowing textual tokens to iteratively query audio features, thus enhancing fusion expressiveness.

Although attention-based fusion strategies are effective, they require task-specific queries to adapt to dataset variations, limiting unified multimodal modeling. To address this, \cite{articlenew22, articlenew23} proposed a bidirectional cross-attention mechanism to improve adaptability and generalization across datasets. However, while bidirectional cross-attention has proven effective, it may introduce noise when dealing with emotionally inconsistent modalities, leading to model confusion and performance degradation.

\section{APPROACH}

\subsection{Problem Statement} 
Our proposed model takes raw speech and text as input, aiming to integrate acoustic information from speech and semantic information from text to comprehensively determine the conveyed emotion. The input speech and text are first processed by their respective encoders, resulting in speech sequence vectors \( S_{seq} = \{s_1, s_2, \dots, s_m\} \) and text sequence vectors \( T_{seq} = \{t_1, t_2, \dots, t_n\} \), where \( m \) and \( n \) denote the lengths of the encoded sequences. The outputs of the encoders are optimized through the \textbf{ORL Module}.
The optimized representations are then fed into the \textbf{KF Module}, which outputs the final emotion probability vector  \( P \in \{p_1, p_2, \dots, p_n\} \).

\subsection{Model Architecture} \label{sec:model_architecture}
Fig.~\ref{fig:fig2} illustrates the proposed DRKF. It consists of two key components: the ORL Module, and the KF Module.

\textbf{(1) The ORL Module} comprises three components: Modality Encoding (ME), Progressive Augmentation (PA), and Decoupled Representations (DR). The ME integrates an acoustic encoder and a semantic encoder to extract modality-specific embeddings. The acoustic encoder, based on the pre-trained wav2vec2 model\footnote{\url{https://huggingface.co/audeering/wav2vec2-large-robust-12-ft-emotion-msp-dim}}, transforms raw audio into acoustic embeddings, while the semantic encoder, leveraging the pre-trained RoBERTa model\footnote{\url{https://huggingface.co/FacebookAI/roberta-large}}, encodes raw text into semantic embeddings.
The PE employs two identically structured residual autoencoder networks, each consisting of five residual autoencoder blocks with six linear layers per block. The purpose of this component is to learn the optimal feature augmentation for the acoustic and semantic modalities. The DR  includes contrastive training and emotion modality (Emodality) shuffling mechanisms. The contrastive training method eliminates task-irrelevant modality noise and facilitates the learning of decoupled representations. The Emodality shuffling mechanism restructures optimized modality pairs for downstream processing.

\textbf{(2) The KF Module} comprises three components: the Fusion Encoder (FE), Emotion Classification Submodule (EC) and the Emotion Discrimination Submodule (ED). The FE is a lightweight, self-attention-based encoder that identifies the dominant modality and integrates supplementary emotional information from other modalities. To address potential errors caused by incorrect dominant modality selection under emotionally inconsistent conditions, the ED enforces the fused representation to retain discriminative cues related to intermodal emotional discrepancies. This mechanism ensures that, even when the FE fails to select the optimal modality, the EC can still make accurate predictions by leveraging the preserved inconsistency information. Both ED and EC are implemented as two independent multilayer perceptrons (MLPs).

\subsection{The ORL Module}
\label{sec:ORL Module}
\subsubsection{The Decoupled Representations}
\begin{figure}[ht]
    \centering
    \includegraphics[width=\linewidth]{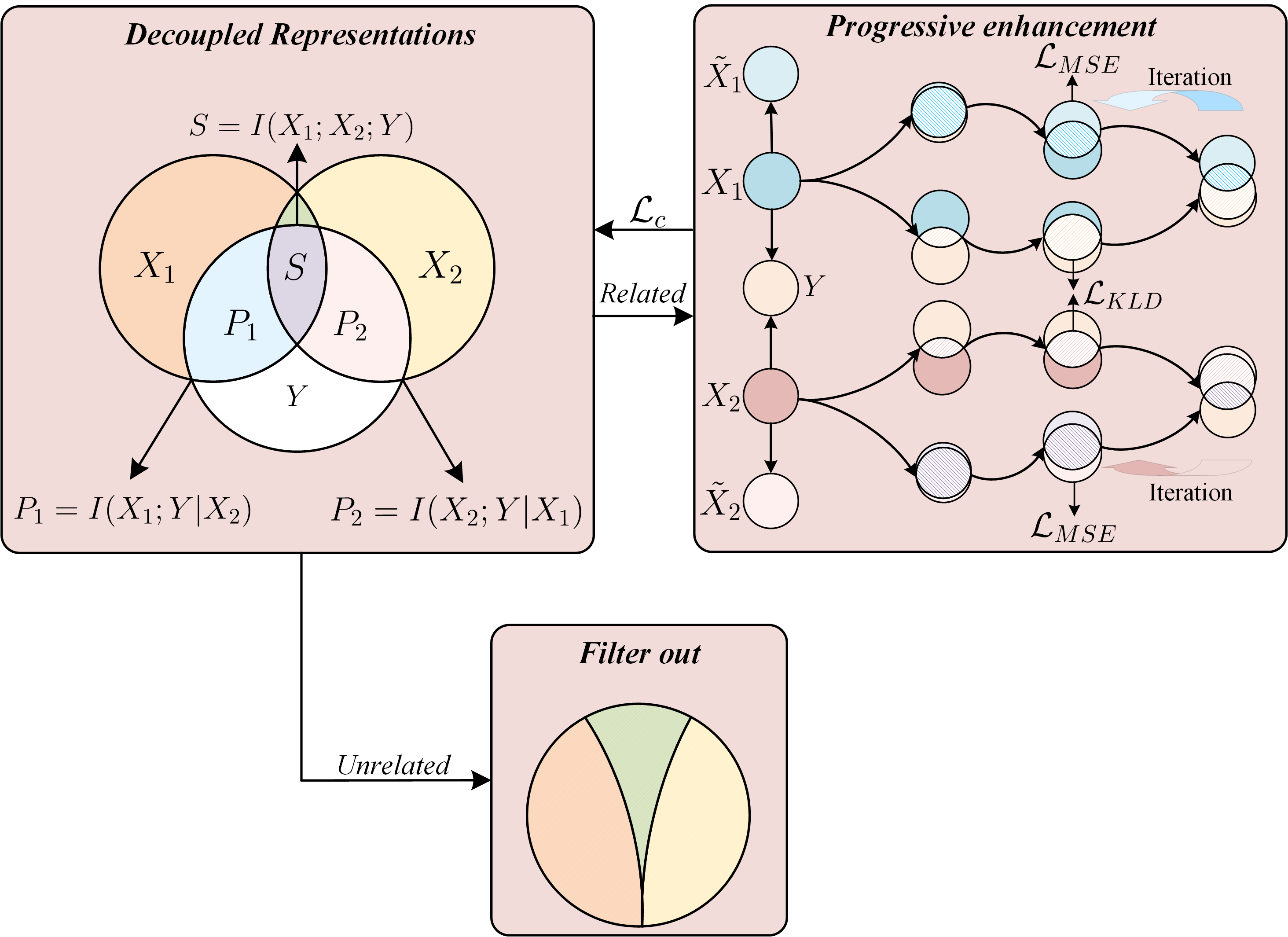}  
    \caption{Decoupled Representations Learning Flowchart.}
    \Description{Decoupled Representations Learning Flowchart.}
    \label{fig:fig3}
\end{figure}

We employed an information-theoretic-based decoupled representation approach to filter out task-irrelevant modality information and optimize task-relevant representations, including modality-shared information \(S\) and modality-specific information \(P\). The task-relevant modality information can be expressed by the following formula: 
\begin{equation}
\label{eq1}
\scalebox{1}{$
I(X_1, X_2; Y) = S + P_1 + P_2
$}
\end{equation}

Where, $I(X_1, X_2; Y)$ is the mutual information between the task variable \( Y \) and the modalities \( X_1 \) and \( X_2 \).
\begin{equation}
\label{eq2}
S = I(X_1; X_2) - I(X_1; X_2 \mid Y)
\end{equation}
\begin{equation}
\label{eq3}
P_1 = I(X_1; Y \mid X_2), \quad P_2 = I(X_2; Y \mid X_1)
\end{equation}

Where, $S$ is the task-relevant modality-shared mutual information, $P_1$ and $P_2$ are the task-relevant modality-specific mutual information within each modality.
\begin{equation}
\label{eq4}
I(X_1; X_2) = \int p(x_1, x_2) \log \frac{p(x_1, x_2)}{p(x_1)p(x_2)} \, dx_1 \, dx_2
\end{equation}
\begin{equation}
\label{eq5}
\scalebox{0.95}{
$I(X_1; X_2 \mid Y) = \int p(x_1, x_2, y) \log \frac{p(x_1, x_2 \mid y)}{p(x_1 \mid y)p(x_2 \mid y)} \, dx_1 \, dx_2 \, dy$}
\end{equation}

Where, $I(X_1; X_2)$ represents the total mutual information between the two modalities $X_1$ and $X_2$, \( I(X_1; X_2 \mid Y) \) represents the conditional mutual information between \( X_1 \) and \( X_2 \) given the task \( Y \), reflecting task-irrelevant modality-shared mutual information.

Calculating mutual information, as shown in the formula above, requires a closed-form density function and a log-density ratio between the joint and marginal distributions in a manageable form. However, in real-world machine learning tasks, we only have access to samples from the joint distribution, making direct computation of mutual information difficult and forcing us to rely on approximation methods. 

\subsubsection{The Progressive Augmentation}

To address the challenge of directly computing mutual information, we introduce contrastive mutual information estimation, formulated as follows: 

\begin{equation} 
\label{eq14} 
I\left(X_{1} ; X_{2}\right) = \mathbb{E}{x_{1}, x_{2}, x_{2}^{-}}\left[\log \frac{\exp f\left(x_{1}, x_{2}\right)}{\sum_{M} \exp f\left(x_{1}, x_{2}^{-}\right)}\right] 
\end{equation}

\begin{equation} 
\label{eq15}
\scalebox{1.0}{
$I\left(X_{1} ; X_{2} \mid Y\right) = 
\mathbb{E}_{y, x_{1}, x_{2}, x_{2}^{-}} 
\left[
\log 
\frac{
    \exp f\left(x_{1}, x_{2}, y\right)
}{
    \sum_{M} 
    \exp f\left(x_{1}, x_{2}^{-}, y\right)
}
\right]$}
\end{equation}

In the above equations, \( M \) represents the batch size during training. The expectation operator \( \mathbb{E} \) is taken over the joint distribution of positive and negative sample pairs, with or without conditioning on the label \( y \). The function \( f(\cdot, \cdot) \) measures \text{correlation} between inputs. Positive samples \( (x_1, x_2) \) or \( (x_1, x_2, y) \) share semantic alignment, while negative samples \( x_2^{-} \) or \( (x_2^{-}, y) \) are drawn from other instances in the batch.

Although contrastive mutual information estimation circumvents the challenges of direct computation, it still faces a critical limitation: the representation gap between modalities and task labels further increases the difficulty of mutual information estimation. As shown in Eq.~\eqref{eq15}, \( x_1 \) and \( x_2 \) follow continuous distributions in their respective feature spaces, whereas \( y \) is discrete. This modality-label distribution mismatch makes score estimation more challenging. To tackle these challenges, we propose a progressive modality augmentation strategy that guides the alignment between modalities and between modality and label distributions by iteratively learning the optimal augmented modality. The optimal augmented modality is defined as follows:

\begin{itemize}
 \item Optimal augmented modality: 
 When \(  I(X; Y) = I(X; \tilde{X}_1
) \), \( \tilde{X}_1
 \) is the optimal unimodal augmentation of \( X \), which implies that the only information shared between \( X \) and \( \tilde{X}_1
 \) is task-relevant, and that \( X \) and \( \tilde{X}_1
 \) lie within the same subspace.
\end{itemize}

Our proposed progressive augmentation strategy, as illustrated in Fig.~\ref{fig:fig3}, is designed to learn the optimal augmented modality. Unlike traditional static feature augmentation methods, our proposed progressive augmentation strategy is a dynamic optimization approach based on original modality features, enabling a stepwise optimization process to achieve optimal modality augmentation. To guide the learning of the augmented features, we introduce two carefully designed constraints. First, by constraining the distribution discrepancy between the augmented modality and the original modality, we ensure that the augmented features reside in the same subspace as the original modality features, thus reducing modality heterogeneity. Second, by minimizing the discrepancy between the augmented modality and the task label distribution, we ensure that the augmented features are aligned with the task label distribution to the greatest extent. Finally, we control the model's overall performance by adjusting the weights of these two constraints.

The process of progressive augmentation can be described by the following equations:

We utilize a residual autoencoder $f_{AE}$ to learn the optimal unimodal augmentation $\tilde{T}_{seq}, \tilde{T}_{cls}$ for the text feature sequence.
  \begin{equation}
  \label{eq9}
    \tilde{T}_{seq} = f_{AE}(T_{seq}), \quad \tilde{T}_{cls} = \mathrm{avg}(\tilde{T}_{seq})
  \end{equation}

Where $T_{seq} \in \mathbb{R}^{d_n \times d_z}$ denotes the feature sequence output of the text encoder, with $d_n$ as the sequence length and $d_z$ as the dimensionality of each sequence element. $\tilde{T}_{cls} \in \mathbb{R}^{d_z}$ represents the global feature vector obtained by average pooling the sequence feature vector.

Similarly, the optimal unimodal augmentation for the speech feature sequence is calculated as follows:
  \begin{equation}
  \label{eq10}
     \tilde{S}_{seq} = f_{AE}(S_{seq}), \quad \tilde{S}_{cls} = \mathrm{avg}(\tilde{S}_{seq})
   \end{equation}

Where $S_{seq} \in \mathbb{R}^{d_n \times d_z}$ denotes the feature sequence output of the speech encoder, with $d_n$ as the sequence length and $d_z$ as the dimensionality of each sequence element. $\tilde{S}_{cls} \in \mathbb{R}^{d_z}$ represents the global feature vector obtained by average pooling the sequence feature vector.

To ensure that the augmented modality remains in the same subspace as the original modality, we use the Mean Squared Error (MSE) loss to constrain the learning space of the augmented modality. Meanwhile, the Kullback-Leibler divergence (KLD) loss is employed to enforce the augmented modality to learn the distribution of the task labels. The calculation formula is as follows:
\begin{equation}
\label{eq11}
    \mathcal{L}_{MSE} = \frac{1}{2M} \sum_{i=1}^{M} \left[ \left\| S_{seq}^i - \tilde{S}_{seq}^{i} \right\|^2 + \left\| T_{seq}^i - \tilde{T}_{seq}^{i} \right\|^2 \right]
\end{equation}
\begin{equation} 
\label{eq12}
    \mathcal{L}_{KLD} = \frac{1}{M} \sum_{i=1}^{M} \sum_{c=1}^{C} y_{i}^{c} \log \frac{y_{i}^{c}}{\hat{y}_{i}^{c}}
\end{equation}

\begin{equation}
\label{eq13}
    \mathcal{L}_{a} = \alpha \cdot \mathcal{L}_{MSE} +  \mathcal{L}_{KLD}
\end{equation}

Here, $S_{seq}^i$ and $T_{seq}^i$ represent the original feature sequence vectors of the $i$-th sample in the speech and text modalities, respectively. 
$\tilde{S}_{seq}^{i}$ and $\tilde{T}_{seq}^{i}$ represent the augmented feature sequence vectors. $M$ is the batch size, and $C$ denotes the total number of emotion categories. The true label of the \(i\)-th sample in the fusion modality for category \(c\) is denoted as \(y_i^c\), while \(\hat{y}_i^c\) represents the predicted probability for the same sample and category. \(\mathcal{L}_a\) refers to the final augmentation loss, and \(\alpha\) is a hyperparameter.

\subsubsection{Contrastive Mutual Information Estimation}

The maximum task-relevant modality mutual information can be transformed into minimizing the negative of contrastive mutual information. Therefore, the objective function for our decoupled representation learning method is defined as follows:
\begin{equation}
\label{eq17}
\scalebox{0.94}{
$z_S^i = g(S_{cls}^i),  {\tilde{z}_S^i} = g({\tilde{S}_{cls}^i}), z_T^i = g(T_{cls}^i),  {\tilde{z}_T^i} = g({\tilde{T}_{cls}^i})$}
\end{equation}
\begin{equation}
\label{eq18}
\mathcal{L}_{MI_{S_i}} = -\log \frac{\exp(\text{sim}(z_S^i, {\tilde{z}_S^i}) / \tau)}{\sum_{k=1}^{2M} [k \neq i] \exp(\text{sim}(z_S^i, \tilde{z}_S^k) / \tau)}
\end{equation}
\begin{equation}
\label{eq19}
\mathcal{L}_{MI_{T_i}} = -\log \frac{\exp(\text{sim}(z_T^i, {\tilde{z}_T^i}) / \tau)}{\sum_{k=1}^{2M} [k \neq i] \exp(\text{sim}(z_T^i, \tilde{z}_T^k) / \tau)}
\end{equation}
\begin{equation}
\label{eq20}
\mathcal{L}_{MI_{S_i \& T_i}} = -\log \frac{\exp(\text{sim}(z_S^i, z_T^i) / \tau)}{\sum_{k=1}^{M} \exp(\text{sim}(z_S^i, z_T^k) / \tau)}
\end{equation}

In the above formulas, $g(\cdot)$ represents a projection function implemented via an MLP; $\text{sim}(\cdot, \cdot)$ represents the cosine similarity; $\mathcal{L}_{MI_{S_i}}$ and $\mathcal{L}_{MI_{T_i}}$ denote the intra-modality CMIE objective functions, while $\mathcal{L}_{MI_{S_i \& T_i}}$ corresponds to the intermodality CMIE objective function.

Finally, our CMIE optimization objective $\mathcal{L}_c$ is defined as follows:
\begin{equation}
\label{eq21}
\mathcal{L}_c = \frac{1}{M} \sum_{i=1}^{M} \left( \mathcal{L}_{MI_{S_i}} + \mathcal{L}_{MI_{T_i}} +\mathcal{L}_{MI_{S_i \& T_i}} \right)
\end{equation}
\subsection{The KF Module} 
\label{sec:knowledge_fusion}
\subsubsection{The Fusion Encoder} 
To enable the proposed knowledge fusion module to effectively model the complex interactions between different modalities, we applied a specialized concatenation process to the multimodal input sequences, as shown in Equation~\ref{eq22}.
\begin{equation}
\label{eq22}
X_{fusion} = f_{FE} \left( \text{Concat}\left( C_{cls}, S_{seq}, C_{sep}, T_{seq}, C_{sep} \right) \right)
\end{equation}

Where, $Concat(\cdot)$ denotes the concatenation function, $ C_{cls} $  and $C_{sep}$ represent the classification token vector and separator token vector, respectively. $ C_{cls} $ is designed to aggregate global information from the input features during the modality fusion process, while $C_{sep}$ acts as a boundary to distinguish between the two modalities, facilitating the ED in learning modality-consistent information. Finally, $f_{FE}(\cdot)$ processes the concatenated sequence to generate the fused feature vector $X_{fusion}$.

\subsubsection{The Emotion Discrimination Submodule} 
The ED is implemented as an MLP, trained on data generated through the Emodality shuffling process in the ORL module. Specifically, for each batch of data, samples from different modalities are randomly combined to create new speech-text pairs. If the original emotion labels of both modalities in a newly formed pair match, the emotional information is considered consistent (assigned a label of 1). Conversely, if the labels do not match, the emotional information is deemed inconsistent (assigned a label of 0). The optimization function for this module is as follows:

\begin{equation}
\label{eq24}
\mathcal{L}_b = -\frac{1}{M^2} \sum_{i=1}^{M^2} \left[ Y_{i} \log \hat{Y}_{i} + (1 - Y_{i}) \log (1 - \hat{Y}_{i}) \right]
\end{equation}
Where, $M$ is the batch size, $Y_{i}$ is the true label of the $i$-th sample in the binary emotion decoupling task, and $\hat{Y}_{i}$ represents the predicted probability of the $i$-th sample output.

\subsubsection{The Emotion Classification Submodule} 
\label{sec:Emotion Classification and Discrimination}
We employ an independent MLP to perform multimodal emotion classification.  Specifically, the EC takes as input the correctly matched sample pairs generated by the Emodality shuffling process. The calculation processes for the emotion classification loss $\mathcal{L}_f$ is presented in Equation~\ref{eq23}.
\begin{equation}
\label{eq23}
\mathcal{L}_f = -\frac{1}{M} \sum_{i=1}^{M} \sum_{c=1}^{C} y_{i}^{c} \log \hat{y}_{i}^{c}
\end{equation}

\begin{table*}[ht]
\centering
\caption{Model Performance Comparison on the IEMOCAP Dataset}
\begin{tabular*}{0.9\textwidth}{@{\extracolsep{\fill}}lcccc}
\toprule
\textbf{Models} & \textbf{Audio \& Text Encoder} & \textbf{ACC (\%)} & \textbf{WACC (\%)} & \textbf{Avg (\%)} \\
\midrule
GBAN \cite{liu2022group}             & CNN-LSTM     & 70.1   & 72.4        & 71.2      \\
MSER-MVAM \cite{feng2023multimodal}  & CNN-LSTM     & 74.2   & 75.4        & 74.8      \\
MSER-CADF \cite{khan2024mser}        & CNN-GRU      & 77.2   & 76.5        & 76.8       \\
MCFN \cite{zhang2023dual}            & CNN-Roberta  & 77.8   & 76.0        & 76.9      \\
SAMS \cite{hou2023semantic}          & BiGRU-Bert   & 78.1   & 76.6        & 77.3       \\
LLMSER \cite{santoso2024large}       & BiLSTM-Bert  & 78.3   & \underline{78.1}        & 78.2      \\
KS-Transformer \cite{article37}      & Wav2vec-Roberta & 75.3  & 74.3      & 74.8       \\
KBCAM \cite{zhao2023knowledge}       & Wav2vec2-Bert & 77.0    & 75.5      & 76.2      \\
DBT \cite{yi2023dbt}                 & Wav2vec2-Roberta & \underline{78.9} & 77.8 & \underline{78.3}    \\
\textbf{Ours(ORKF)}      &  Wav2vec2-Roberta  & \textbf{80.7} & \textbf{79.9} & \textbf{80.3}\\
\textbf{\(\Delta\)Sota}
    & ---  & \textbf{\(\uparrow\) 2.28} & \textbf{\(\uparrow\)2.30} & \textbf{\(\uparrow\)2.55} \\
\bottomrule
\end{tabular*}
\label{tab:iemocap_performance}
\begin{flushleft}
    \textbf{Notes:}  
    \textbf{Bold values indicate the best performance.}  
    \underline{Underlined values} denote the second-best performance.  
    \(\Delta\)Sota represents the relative improvement of our proposed method compared to the second-best model.
    (\(\uparrow\)) indicates an improvement over the second-best performance, where higher values are better. (\(\downarrow\)) indicates a decrease relative to the best performance, where lower values are better.
\end{flushleft}
\end{table*}

Where, $M$ is the batch size, $C$ is the number of emotion categories, $y_{i}^{c}$ is the true label of the $i$-th sample in the fusion modality for category $c$, and $\hat{y}_{i}^{c}$ represents the predicted probability of the $i$-th sample in the fusion modality for category $c$.

Finally, our objective function, denoted as $\mathcal{L}$, is formally defined in Equation~\ref{eq25}.
\begin{equation}
\label{eq25}
\mathcal{L} = \mathcal{L}_{a} + \beta \cdot \mathcal{L}_{c} + \gamma \cdot \mathcal{L}_{f} + \delta \cdot \mathcal{L}_{b}
\end{equation}

Where $\beta$, $\gamma$, and $\delta$ are hyperparameters.
\section{Experiment Settings}

\subsection{Datasets}

We validated our proposed model on three publicly available multimodal datasets, IEMOCAP \cite{article52}, MELD \cite{article53} and M3ED \cite{m3ed}. Specifically, IEMOCAP is a recorded dialogue dataset with emotion labels including anger, happiness, sadness, frustration, excitement, fear, surprise, disgust, and others. To ensure consistency with previous research, we focused on four emotion categories: happiness, sadness, anger, and neutral, where excitement was merged into the happiness category, resulting in a total of 5,531 samples. The experiments followed a five-fold leave-one-session-out strategy, Using Unweighted Accuracy (ACC), Weighted Accuracy (WACC), and their average (Avg) to evaluate model performance. MELD is a challenging multi-party conversation dataset, annotated with seven emotion labels. Unlike IEMOCAP, this dataset is divided into training, development, and test sets, providing a standardized training and evaluation strategy for models. WACC, Weighted F1 score (WF1) and Avg were used to assess the performance of the models on this dataset. M3ED is the first Chinese multi-label emotion dialogue dataset. The utterance-level emotion labels include seven categories: happiness, surprise, sadness, disgust, anger, fear, and neutral. Following previous studies, we used Precision, Recall, ACC, Micro-F1 (F1) score and the Avg as evaluation metrics to assess model performance.

\subsection{Implementation Details}
Our model is implemented using the PyTorch framework, with AdamW as the optimizer, a learning rate of 1e-5, and a batch size of 4. The output dimension of the projection head function in contrastive learning is set to 1024, and the multimodal fusion layer has 8 attention heads. The values of the loss function hyperparameters are set to 0.2, 0.2, 1.0, and 0.2, respectively. Our training is conducted on a Linux system with an A100 GPU, for a total of 100 epochs.

\subsection{Baseline Models}
To validate the effectiveness of the proposed method, we compared ORKF with the current advanced baseline methods. The baselines used to evaluate ORKF across different datasets are as follows. It is important to note that some baseline models are evaluated on multiple datasets, and we provide their descriptions only when they are mentioned for the first time.

Compared Methods for IEMOCAP Dataset: GBAN \cite{liu2022group} with gated attention fusion; DIMMN \cite{wen2023dynamic} using dynamic memory interaction; MSER-CADF \cite{khan2024mser} with cross-attention fusion; MCFN \cite{zhang2023dual} employing dual-stream temporal-spatial modeling; SAMS \cite{hou2023semantic} aligning semantics across modalities; LLMSER \cite{santoso2024large} enhancing prompts in language models; KS-Transformer \cite{article37} using pre-trained feature extraction and early fusion; KBCAM \cite{zhao2023knowledge} incorporating Bayesian attention with external knowledge; DBT \cite{yi2023dbt} utilizing dual-branch Transformer with fine-tuning fusion. Compared Methods for MELD Dataset: MCSCAN \cite{sun2021multimodal} with parallel cross/self-attention; DIMMN \cite{wen2023dynamic} with dynamic memory integration; SACMA \cite{guo2024speaker} integrating speaker-aware emotion recognition; SMCN \cite{hou2022multi} self-guided modality alignment; RMERCT \cite{xie2021robust} using Transformer-based cross-modal fusion; SMIN \cite{lian2022smin} semi-supervised multimodal learning; HiMul-LGG \cite{fu2024himul} hierarchical decision fusion strategy. Compared Methods for M3ED Dataset: MSCNN-SPU \cite{MSCNN-SPU} integrating multi-scale CNN with statistical pooling; M-TLEAF \cite{M-TLEAF} using bidirectional GRU and Transformer fusion; CARAT \cite{carat} employing contrastive feature reconstruction and aggregation.

\subsection{Results}
\begin{table*}[ht]
\centering
\caption{Model Performance Comparison on the MELD Dataset}
\begin{tabular*}{0.9\textwidth}{@{\extracolsep{\fill}}lcccc}
\toprule
\textbf{Models} & \textbf{Audio \& Text Encoder} & \textbf{WACC  (\%)} & \textbf{WF1  (\%)} & \textbf{Avg  (\%)} \\ \midrule
MCSCAN \cite{sun2021multimodal}      & CNN \& LSTM-LSTM   & N/A     & 59.2      & 59.2\textsuperscript{\ddag}   \\
DIMMN \cite{wen2023dynamic}          & Attention-CNN    & 60.6    & 58.6      & 59.6      \\
SACMA \cite{guo2024speaker}          & LSTM-TextCNN     & 62.3    & 59.3      & 60.8    \\
MCFN \cite{zhang2023dual}            & CNN-Roberta      & 64.5    & 62.2      & 63.3    \\
SMCN \cite{hou2022multi}             & GRU-Bert         & 64.9    & 62.3      & 63.6    \\
SAMS \cite{hou2023semantic}          & BiGRU-Bert       & 65.4    & 62.6      & 64.0   \\
RMERCT \cite{xie2021robust}          & WaveRNN-GPT      & 63.1    & 64.0      & 63.5    \\
SMIN \cite{lian2022smin}             & Wav2vec-Roberta  & 65.5    & 64.5      & 65.0    \\
HiMul-LGG \cite{fu2024himul}         & BiGRU-Roberta    & \underline{66.2}    & \underline{65.1}      & \underline{65.6}  \\
\textbf{Ours(ORKF)}                   & Wav2vec2-Roberta  & \textbf{66.7} & \textbf{65.4} & \textbf{66.0}\\
\textbf{\(\Delta\)Sota}
    & ---  & \textbf{\(\uparrow\) 0.75} & \textbf{\(\uparrow\)0.46} & \textbf{\(\uparrow\)0.60} \\
\bottomrule
\end{tabular*}
    \label{tab:meld_comparsion}
\begin{flushleft}
    \textbf{Notes:} 
    N/A indicates that the metric value was not provided in the original paper.
    \textsuperscript{\ddag} denotes that the average value was calculated from known experimental results due to irreproducibility. 
\end{flushleft}
\end{table*}

\begin{table*}[ht]
\centering
\caption{Model Performance Comparison on the M3ED Dataset}
\resizebox{0.9\textwidth}{!}{
\begin{tabular}{lcccccc}
\toprule
\textbf{Models} & \textbf{Audio \& Text Encoder} & \textbf{Precision (\%)} & \textbf{Recall (\%)} & \textbf{ACC (\%)} & \textbf{F1 (\%)} & \textbf{Avg (\%)} \\ \midrule
MSCNN-SPU \textsuperscript{\dag} \cite{MSCNN-SPU} & CNN-TextCNN   & 44.3 & 50.1 & 45.0 & 47.0 & 46.6 \\
M-TLEAF \textsuperscript{\dag} \cite{M-TLEAF} & CNN-BERT          & 45.2 & 49.1 & 46.7  & 47.1 & 47.0 \\
MCFN \textsuperscript{\dag} \cite{zhang2023dual} & CNN-Roberta    & 44.7 & 50.9 & 46.1 & 47.6 & 47.3 \\
SAMS \textsuperscript{\dag} \cite{hou2023semantic} & BiGRU-Bert   & \underline{47.1} & \underline{51.5} & \textbf{51.1} & \underline{49.2} & \underline{49.7} \\
CARAT\textsuperscript{\dag} \cite{carat} & Transformer Encoder-Based & 45.0 & 51.4 & 44.3 & 48.0 & 47.1\\
\textbf{Ours(ORKF)}  & Wav2vec2-Roberta         & \textbf{52.6} & \textbf{51.6} & \underline{50.6} & \textbf{52.0} & \textbf{51.7}\\
\textbf{\(\Delta\)Sota}
    & ---  & \textbf{\(\uparrow\) 11.6} & \textbf{\(\uparrow\)0.38} & \textbf{\(\downarrow\)0.98} & \textbf{\(\uparrow\)5.69}& \textbf{\(\uparrow\)4.02}\\
\bottomrule
\end{tabular}}
\label{tab:m3ed_comparsion}
\begin{flushleft}
\textbf{Notes:} \textsuperscript{\dag} The results are obtained through our own reproduction experiments.
\end{flushleft}
\end{table*}

As shown in Table~\ref{tab:iemocap_performance}, the proposed ORKF method achieves the best overall performance on the IEMOCAP dataset, with an ACC of 80.7\%, a WACC of 79.9\%, and an average (Avg) of 80.3\%. In terms of ACC and Avg, ORKF achieves relative improvements of approximately 2.28\% and 2.55\%, respectively, compared to the second-best method DBT (ACC of 78.9\%, Avg of 78.3\%). For the WACC metric, ORKF shows a relative improvement of about 2.30\% over the second-best method LLMSER (WACC of 78.1\%).

To further validate the performance of ORKF, we evaluated the model on the MELD dataset. The experimental results are presented in Table~\ref{tab:meld_comparsion}.

According to the experimental comparison results in Table~\ref{tab:meld_comparsion}, ORKF demonstrates superior performance even on the highly imbalanced MELD dataset. Specifically, ORKF achieves a WACC of 66.7, a WF1 of 65.4, and an average score of 66.0, all of which represent the best results among the compared methods. Although ORKF achieved SOTA performance on both the IEMOCAP and MELD datasets, it should be noted that these datasets are English datasets with single-label annotations, where task-relevant information may primarily rely on shared mutual information. To further validate the model's performance, we introduced the Chinese multi-label emotion recognition dataset M3ED for testing.

As shown in Table~\ref{tab:m3ed_comparsion}, even on the more complex multi-label Chinese emotion recognition dataset, ORKF demonstrates strong performance, achieving an F1 score of 52.0 and an average score of 51.7, reaching SOTA results.

Overall, the comparative experimental results demonstrate that ORKF effectively integrates information from multiple modalities, achieving robust emotion recognition.

\subsection{Ablation Study}
To evaluate the effectiveness of the proposed strategy, we conducted ablation experiments on the IEMOCAP, MELD, and M3ED datasets, with the results presented in Table~\ref{tab:strategy}. 

\begin{table*}[ht]
  \centering
  \caption{Results for Strategy Analysis}
  \setlength{\tabcolsep}{10pt} 
  \renewcommand{\arraystretch}{1.3} 
  \begin{tabular*}{0.9\linewidth}{@{\extracolsep{\fill}} cccccc cccccc @{}}
    \toprule
    \multicolumn{6}{c}{Methods} & \multicolumn{2}{c}{IEMOCAP} & \multicolumn{2}{c}{MELD} & \multicolumn{2}{c}{M3ED} \\
    \cmidrule(lr){1-6} \cmidrule(lr){7-8} \cmidrule(lr){9-10} \cmidrule(lr){11-12}
    BME & PCMI & CMIE & $FE_c$ & $FE_s$ & \(ED\) & ACC & WACC & WACC & WF1 & ACC & F1 \\
    \cmidrule(lr){1-12}
    \checkmark &  &  &  &  \checkmark&  & 78.1 & 77.0 & 64.5 & 63.3 & 47.1 & 49.3 \\
    \checkmark &  & \checkmark &  &  \checkmark &   & 78.4 & 77.5 & 64.7 & 63.4 & 47.3 & 49.9 \\
    \checkmark & \checkmark & &  &  \checkmark &  & 79.7 & 78.4 & 65.6 & 64.7 & 47.9 & 50.8\\
    
    \checkmark &  &  &  &  \checkmark & \checkmark & 79.0 & 78.1 & 65.1 & 63.5 &  48.1 & 51.1 \\

    \checkmark & \checkmark &  & \checkmark &  & \checkmark & 78.2 & 76.3 & 64.3 & 63.2 & 46.7 & 48.5 \\ 
    \checkmark & \checkmark &  &  & \checkmark & \checkmark & \textbf{80.7} & \textbf{79.9} & \textbf{66.7} & \textbf{65.4} & \textbf{50.6} & \textbf{52.0} \\
   
    \bottomrule
  \end{tabular*}
  \label{tab:strategy}
    \begin{justify}
    \textbf{Notes:} 
    The \checkmark symbol indicates that the corresponding method is applied. 
    BME refers to the BiModal Encoder. 
    PCMI represents the Progressive Contrastive Mutual Information Estimation.
    CMIE represents the Contrastive Mutual Information Estimation introduced by \cite{article41}.
\textit{FE}\textsubscript{c} denotes the fusion encoder with a bidirectional cross-attention mechanism. 
    \textit{FE}\textsubscript{s} denotes the fusion encoder with a self-attention mechanism.
    \textit{ED} refers to the emotion discrimination submodule.
  \end{justify}
\end{table*}
\begin{figure*}[ht]
    \centering
    \begin{subfigure}[b]{0.32\textwidth}
        \includegraphics[width=\linewidth]{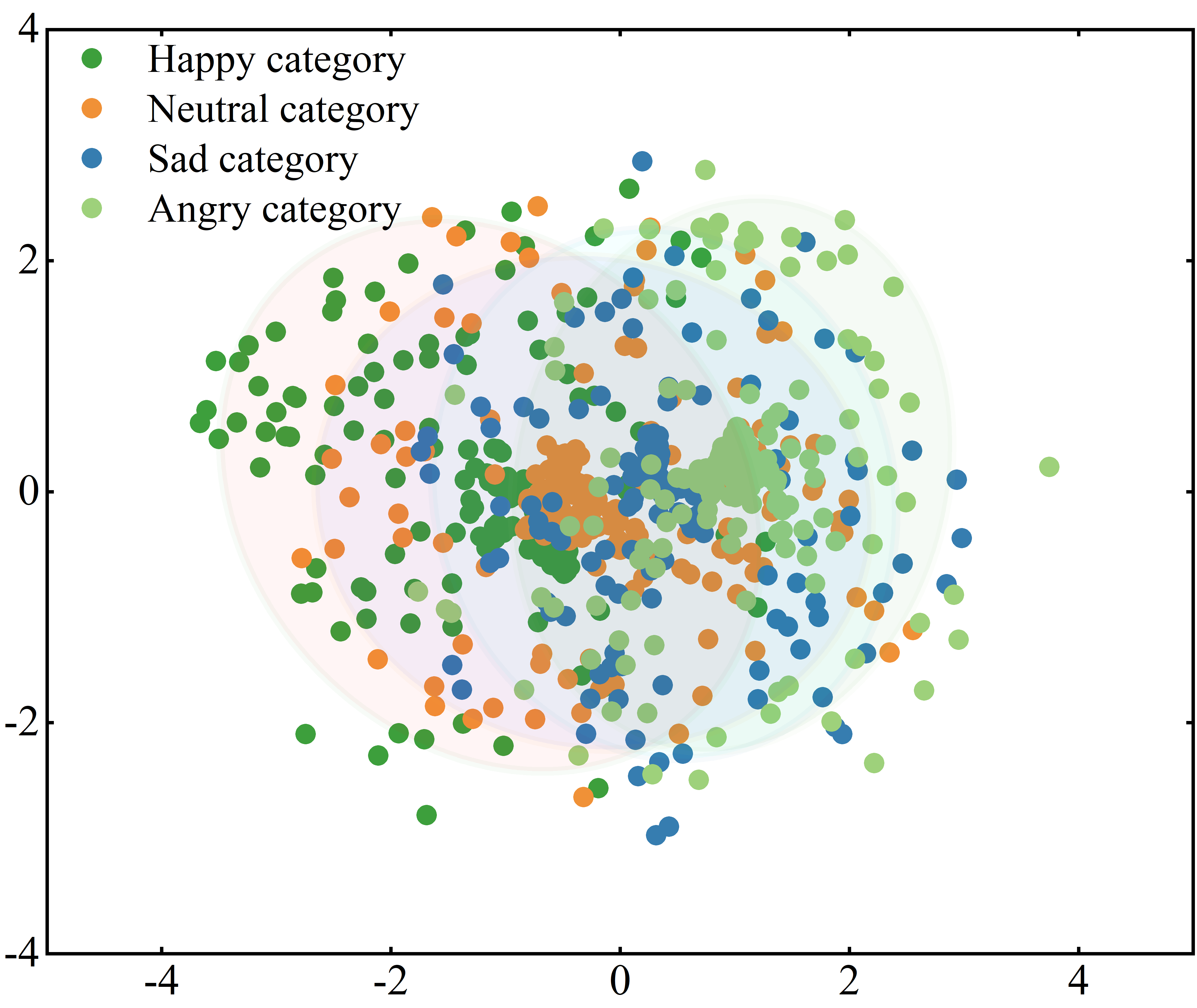}
        \caption{Distribution of Different Emotion Categories before Using PCMI Method}
        \label{fig:fig6a}
    \end{subfigure}
    \hfill
    \begin{subfigure}[b]{0.32\textwidth}
        \includegraphics[width=\linewidth]{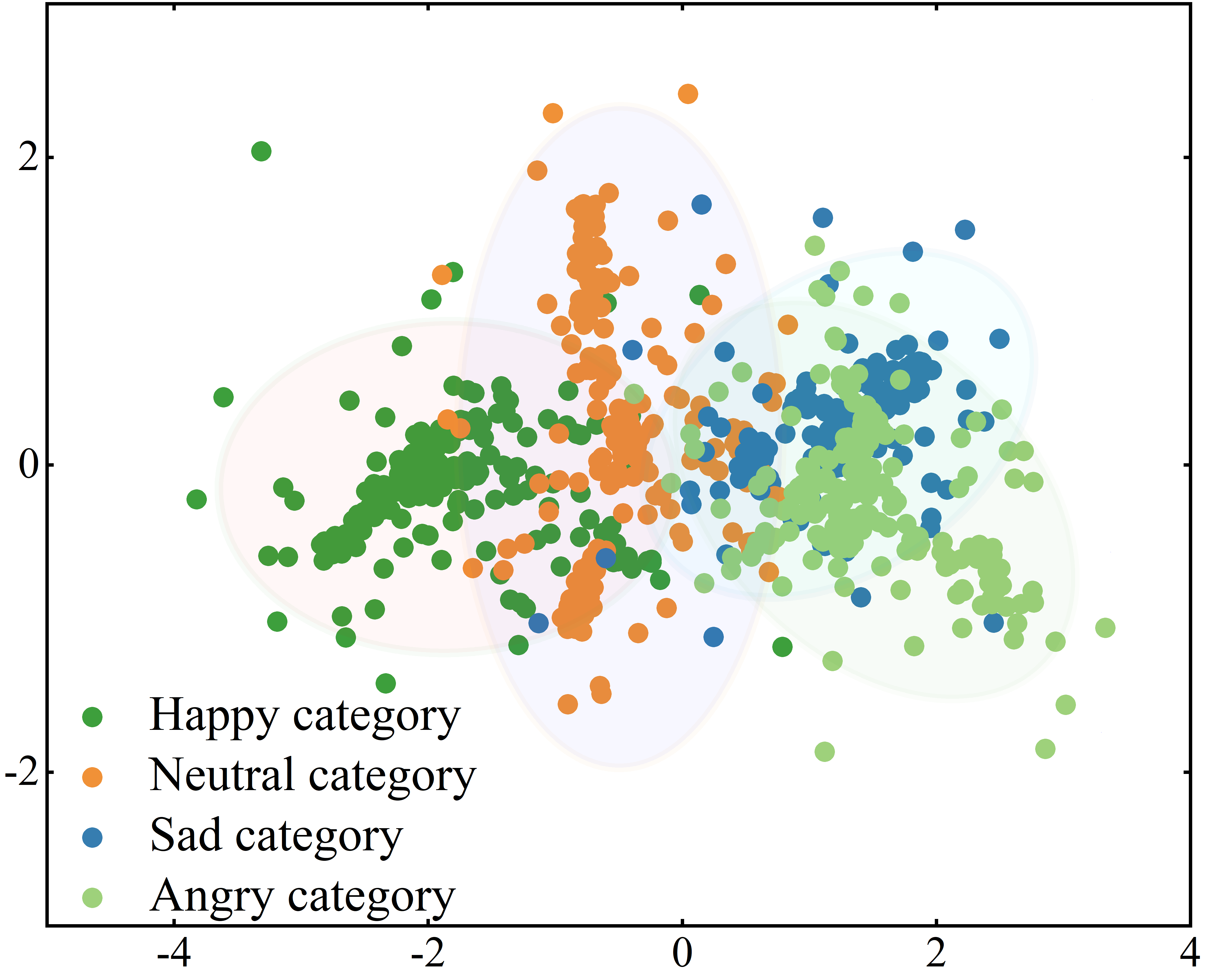}
        \caption{Distribution of Different Emotion Categories after Using CMIE}
        \label{fig:fig6b}
    \end{subfigure}
    \hfill
    \begin{subfigure}[b]{0.32\textwidth}
        \includegraphics[width=\linewidth]{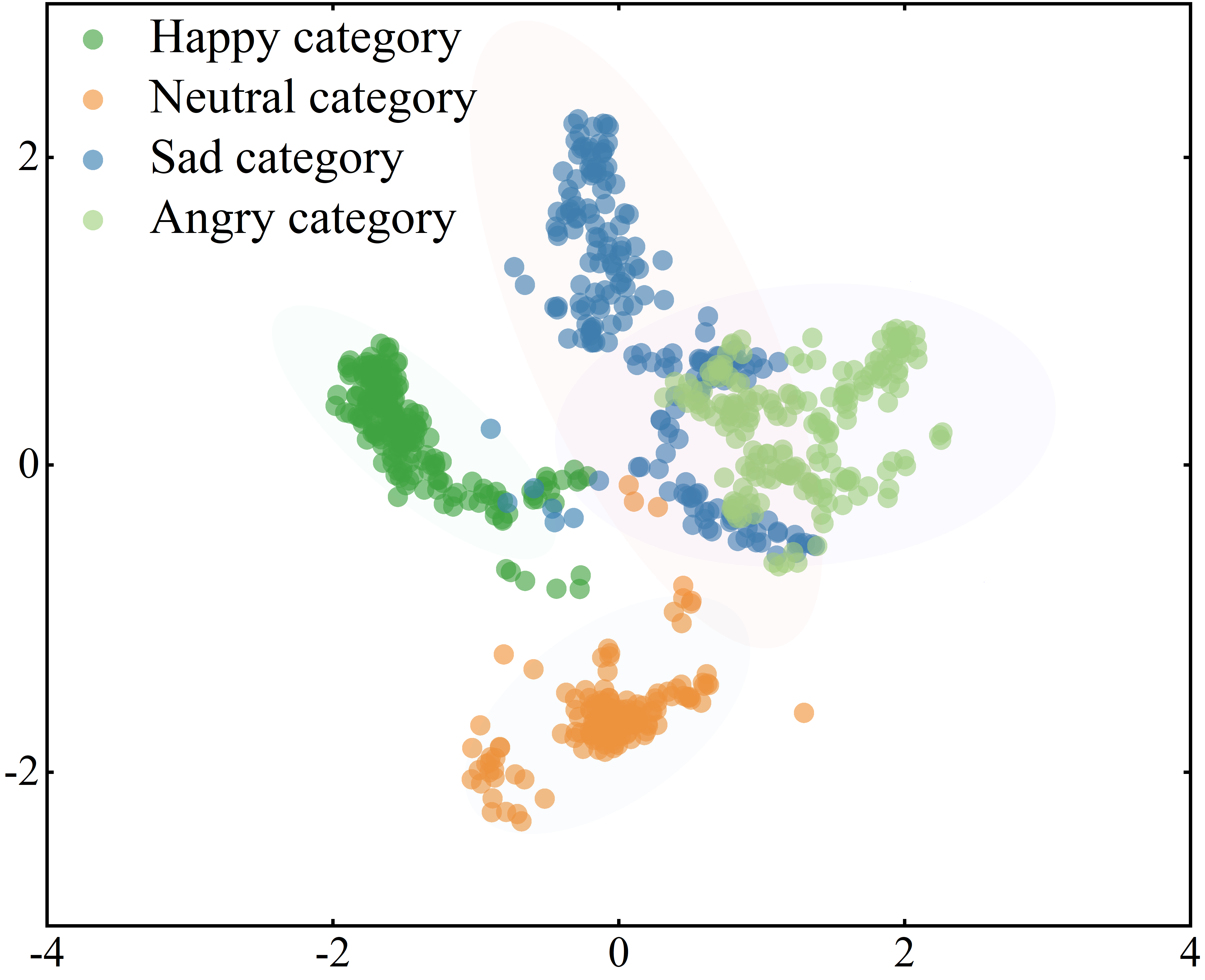}
        \caption{Distribution of Different Emotion Categories after Using PCMI Method}
        \label{fig:fig6c}
    \end{subfigure}
    \caption{Comparison of Emotion Category Distributions: PCMI vs. CMIE and Baseline.}
    \Description{omparison of Emotion Category Distributions: PCMI vs. CMIE and Baseline.}
    \label{fig:fig6}
\end{figure*}
As shown in Table~\ref{tab:strategy}, the comparison between the first and fourth rows demonstrates that removing the ED leads to a noticeable performance drop across all three datasets, highlighting its critical role in helping the fusion module learn effective joint representations. The comparison between the second and third rows further validates the effectiveness of the proposed Progressive Contrastive Mutual Information Estimation (PCMI) approach in enhancing decoupled representation learning and improving overall model performance. 

To further verify the effectiveness of the proposed decoupled representation learning strategy based on PCMI, we visualized the learned embeddings on the test set (session 2) of the IEMOCAP dataset using t-distributed Stochastic Neighbor Embedding (t-SNE), as shown in Figs.~\ref{fig:fig6a}–\ref{fig:fig6c}. From the figures, it can be observed that the decoupled representation learning strategy based on PCMI effectively facilitates the learning of well-structured and discriminative representations.

Finally, the comparison between the fifth and sixth rows shows that the proposed self-attention-based fusion encoder outperforms the traditional cross-attention fusion encoder, offering more substantial gains in emotion recognition accuracy.

\section{Conclusion}
To address the prevalent challenges of modality heterogeneity and emotional inconsistency across modalities in MER tasks, we propose a novel framework named Decoupled Representations with Knowledge Fusion (DRKF). The framework consists of two core modules: the Optimized Representation Learning (ORL) module and the Knowledge Fusion (KF) module. Specifically, the ORL module aims to decouple task-relevant modality-shared and modality-specific information while reducing inter-modality heterogeneity, thereby facilitating more effective multimodal fusion. The KF module is designed to learn a fusion representation that is sensitive to emotional discrepancies across modalities, which enhances the model’s robustness in scenarios where emotional cues from different modalities are not aligned. Extensive experiments on three widely used benchmark datasets for multimodal emotion recognition demonstrate that DRKF outperforms several state-of-the-art models across multiple evaluation metrics, exhibiting strong performance and generalization capabilities.

Despite the promising results achieved by the proposed DRKF model on bimodal emotion recognition tasks, certain limitations remain. The current evaluation is limited to the audio-text bimodal setting, and has not yet been extended to trimodal or higher-order multimodal fusion scenarios. In future work, we plan to further explore the adaptability and scalability of DRKF in more complex multimodal input settings, such as those involving video, speech, and text, to better address the demands of real-world multimodal emotion recognition applications.

\begin{acks}
This work was supported by the National Natural Science Foundation of China (Award No. U22B2061) and the Natural Science Foundation of Sichuan, China (Award No. 2024NSFSC0496).
\end{acks}
\balance
\bibliographystyle{ACM-Reference-Format}
\bibliography{main}
\end{document}